\documentclass{elsart3p}

\usepackage{graphicx}
\usepackage{graphics}
\usepackage{amssymb,amsmath}

\begin{document}

\begin{frontmatter}



\title{Effect of $d$-$f$ hybridization on the Josephson current through Eu-chalcogenides}


\author[address1,address2,cor1]{S.~Kawabata},
\author[address3]{Y.~Asano}, 
\author[address4]{Y.~Tanaka},
\author[address5]{S.~Kashiwaya}

\address[address1]{Nanotechnology Research Institute (NRI), National Institute of Advanced Industrial Science and Technology (AIST), Tsukuba, Ibaraki 305-8568, Japan}

\address[address2]{CREST, Japan Science and Technology Corporation (JST), Kawaguchi, Saitama 332-0012, Japan}

\address[address3]{Department of Applied Physics, Hokkaido University, Sapporo, 060-8628, Japan}

\address[address4]{Department of Applied Physics, Nagoya University, Nagoya, 464-8603, Japan}

\address[address5]{Nanoelectronics Research Institute (NeRI), AIST, Tsukuba, Ibaraki 305-8568, Japan}

\corauth[cor1]{E-mail address: s-kawabata@aist.go.jp}

\date{}

\begin{abstract}
A superconducting ring with a $\pi$ junction made from superconductor/ferromagnetic-metal/superconductor (S-FM-S) exhibits a spontaneous current without an external
magnetic field in the ground state. 
Such $\pi$ ring
provides so-called quiet qubit that can be efficiently decoupled from the fluctuation of the external
field. However, the usage of the FM gives rise to strong Ohmic dissipation. Therefore, the realization
of $\pi$ junctions without FM is expected for qubit applications. We theoretically consider the possibility
of the $\pi$ coupling for S/Eu-chalcogenides/S junctions based on the $d$-$f$ Hamiltonian. 
By use of the Green's function method we found that $\pi$ junction can be formed in the case of the finite
$d$-$f$ hybridization between the conduction $d$  and the localized $f$ electrons. 
\end{abstract}

\end{frontmatter}


\section{Introduction}

The interplay between superconductivity and ferromagnetism has been the subject of study for many decades~\cite{rf:Buzdin}.
Recently theoretical and experimental investigations into the properties of superconductor / ferromagnetic-metal / superconductor (S-FM-S) heterostructures~\cite{rf:Buzdin,rf:Golubov}
have seen an upsurge in interest after the experimental observation of 0-$\pi$ transitions in the Josephson current through S-FM-S junction by Ryanzanov et al.~\cite{rf:Ryanzanov} and by Kontos et al. ~\cite{rf:Kontos}. 
In terms of the Josephson relationship $I_J= I_C \sin \phi$, where $\phi$ is the phase difference between the two superconductor layers, a transition from the 0 to $\pi$ states
implies a change in sign of $I_C$ from positive to negative. 
Physically, such a change in sign of $I_C$ is a consequence of a phase change in the pairing wave-function induced in the FM layer due to the proximity effect.
Josephson junctions presenting a negative $I_C$ are usually called $\pi$ junctions and  such behavior has been observed experimentally.

Recently, $quiet$ qubits consisting of a superconducting loop with a S-FM-S $\pi$ junction have been theoretically proposed~\cite{rf:Ioffe,rf:Blatter}.
In the quiet qubits, a quantum two level system (qubits) is spontaneously generated and therefore it is expected to be robust to the decoherence by the fluctuation of the external magnetic field.
From the viewpoint of the quantum dissipation, however, the structure of S-FM-S junctions is inherently identical with S-N-S junctions (N is a normal nonmagnetic metal).
Therefore a gapless quasiparticle excitation in the FM layer is inevitable.
This feature gives a strong Ohmic dissipation~\cite{rf:Zaikin} and the coherence time of S-FM-S quiet qubits is bound to be very short.

On the other hand, as was predicted by Tanaka and Kashiwaya~\cite{rf:Tanaka}, the $\pi$ junction can be formed in Josephson junctions with ferromagnetic insulators (FI).
By following their theory, we have theoretically proposed a superconductor phase-~\cite{rf:Kawabata1} and flux-type qubits~\cite{rf:Kawabata2,rf:Kawabata3,rf:Kawabata4} based on S-FI-S $\pi$ junctions.
Moreover we have showed that the effect of the dissipation due to the quasi-particle excitation on macroscopic quantum tunneling is negligibly small~\cite{rf:Kawabata3}.
However, in  above studies, we have used  a very simple $\delta$-function model as the FI barrier.
Therefore, the correspondence between this toy mode and the actual band structure of FI is still unclear.
In this paper, we will formulate a numerical calculation method for the Josephson current through FI by taking into account the band structure of FI.
Then we will discuss the possibility of the formation of the $\pi$-coupling for the Josephson junction through the Eu chalcogenides.

\section{Spin-filter effect in Eu-chalcogenides}

Recently spin filtering effect are intensively studied by use of the the Eu chalcogenides~\cite{rf:EuO,rf:Nagahama,rf:Santos}.
The Eu chalcogenides stand out among the FIs as ideal Heisenberg ferromagnets, with a high magnetic moment and a large exchange splitting of the conduction band for Eu $d$-electrons.
Utilizing the exchange splitting ($V_\mathrm{ex}^d$) to filter spins, these materials produce a near-fully spin-polarized current when used as a tunnel barrier.
Of the Eu chalcogenides, EuO has the largest $V_\mathrm{ex}^d$ and the highest Curie temperature ($T_\mathrm{Curie} \sim 69$ K for bulk).

In EuO, the large saturation magnetic moment $\mu g J =7 \mu_B$ per Eu${}^{2+}$ originates from the seven unpaired electrons localized at the $4f$ levels in the energy gap between the
valence band and the conduction band, shown schematically in Fig. 1. 
Ferromagnetic order of the $4f$ spins causes exchange splitting of the conduction band, lowering (raising) the spin-up (-down) band symmetrically by $V_\mathrm{ex}^d/2$. 
Thus, free carriers in the conduction $d$ band are spin-polarized. 
A large exchange splitting of 0.54 eV was first determined by measuring the redshift of the absorption edge in single crystals of EuO cooled below $T_\mathrm{Curie}$~\cite{rf:Busch}.

\begin{figure}[t]
\begin{center}
\includegraphics[width=5.7cm]{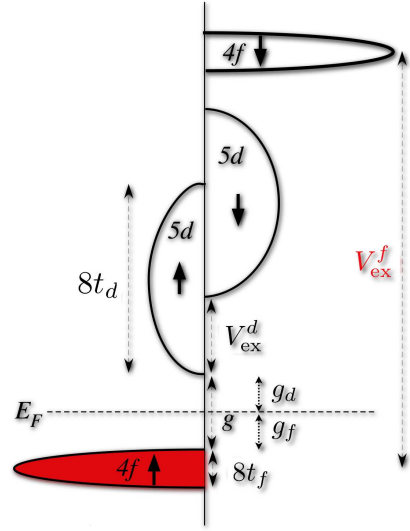}
\end{center}
\caption{ (Color online) The density of states for each spin direction for  the Eu-chalcogenides.
 }
\label{fig1}
\end{figure}

When an ultrathin film of EuO is used as the tunnel barrier between two metallic electrodes, the exchange splitting of the conduction band gives rise to a lower barrier height for spin-up electrons and a
higher barrier height for spin-down electrons. 
Because of the tunnel current depends exponentially on the barrier height~\cite{rf:EuO,rf:Nagahama}, the tunneling probability for spin-up electrons is much greater than for spin-down electrons, leading to a highly spin-polarized current.
This phenomenon is called the spin-filter effect.

\section{Numerical calculation of Josephson current}

In this section, we develop a numerical calculation method of the Josephson current for S-FI-S junctions~\cite{rf:Kawabata-pi,rf:Kawabata-Asano}.
Let us consider the two-dimensional two-band tight-binding model for a S-FI-S junction as shown in Fig.~1.
The vector $\boldsymbol{r}=j{\boldsymbol{x}}
+m{\boldsymbol{y}}$ points to a lattice site, where ${\boldsymbol{x}}$ and ${\boldsymbol{y}}$ are unit vectors in the $x$ and $y$ directions,
respectively.
In the $y$ direction, we apply the periodic boundary condition for the number of lattice sites being $W$.

Electronic states in superconductor are described by the
mean-field Hamiltonian
\begin{align}
 H_{\text{BCS}}=& \frac{1}{2}\sum_{\boldsymbol{r},\boldsymbol{r}^{\prime }}%
\left[ \tilde{c}_{\boldsymbol{r}}^{\dagger }\;\hat{h}_{\boldsymbol{r},%
\boldsymbol{r}^{\prime }}\;\tilde{c}_{\boldsymbol{r}^{\prime }}^{{}}-%
\overline{\tilde{c}_{\boldsymbol{r}}}\;\hat{h}_{\boldsymbol{r},\boldsymbol{r}%
^{\prime }}^{\ast }\;\overline{\tilde{c}_{\boldsymbol{r}^{\prime }}^{\dagger
}}\;\right]   \notag \\
& +\frac{1}{2}\sum_{\boldsymbol{r}\in \text{S}}\left[ \tilde{c}_{%
\boldsymbol{r}}^{\dagger }\;\hat{\Delta}\;\overline{\tilde{c}_{\boldsymbol{r}%
}^{\dagger }}-\overline{\tilde{c}_{\boldsymbol{r}}}\;\hat{\Delta}^{\ast }\;%
\tilde{c}_{\boldsymbol{r}}\right] ,  \label{bcs}
\end{align}%
\begin{align}
\hat{h}_{\boldsymbol{r},\boldsymbol{r}^{\prime }}=& \left[ -t\delta _{|%
\boldsymbol{r}-\boldsymbol{r}^{\prime }|,1}+(-\mu
+4t)\delta _{\boldsymbol{r},\boldsymbol{r}^{\prime }}\right] \hat{\sigma}_{0}
\notag ,
\end{align}%
with 
$\overline{\tilde{c}}_{\boldsymbol{r}}=\left( c_{\boldsymbol{r}%
,\uparrow },c_{\boldsymbol{r},\downarrow }\right) $,
 where
  $
  c_{\boldsymbol{r} ,\sigma }^{\dagger }$ ($c_{\boldsymbol{r},\sigma }^{{}}
$)
 is the creation
(annihilation) operator of an electron at $\boldsymbol{r}$ with spin $\sigma
=$ ( $\uparrow $ or $\downarrow $ ), $\overline{\tilde{c}}$ means the
transpose of $\tilde{c}$,  and $\hat{\sigma}_{0}$ is $2\times 2$ unit matrix. 
In superconductors, the hopping integral $t$ is considered among nearest neighbor sites and we choose $\hat{\Delta}=i\Delta \hat{\sigma}_{2}$, where $\Delta $ is the amplitude 
of the pair potential in the $s$-wave symmetry channel, and $\hat{\sigma}_{2}$ is a Pauli matrix.

\begin{figure}[t]
\begin{center}
\includegraphics[width=8cm]{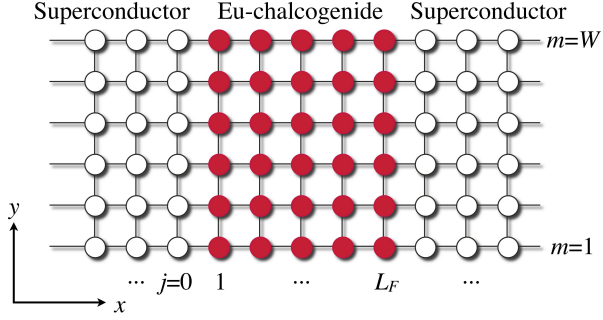}
\end{center}
\caption{ (Color online) A schematic figure of a S/Eu-chalcogenides/S junction on the
tight-binding lattice. 
 }
\label{fig1}
\end{figure}

 In a ferromagnetic insulator, we have used a following two-band Hamiltonian:
 \begin{eqnarray}
H_\mathrm{FI} &=& H_d + H_f + H_{df}, 
\\
H_d &=& 
-t_d \sum_{\boldsymbol{r},\boldsymbol{r}^{\prime },\sigma} 
d_{\boldsymbol{r},\sigma}^\dagger 
d_{\boldsymbol{r}',\sigma}
-\sum_{\boldsymbol{r}} ( 4 t_d -\mu_d)  
d_{\boldsymbol{r},\uparrow}^\dagger 
d_{\boldsymbol{r},\uparrow}
\nonumber\\
&&
- \sum_{\boldsymbol{r}} 
( 4 t_d -\mu_d + V_\mathrm{ex}^d) 
 d_{\boldsymbol{r},\downarrow}^\dagger 
 d_{\boldsymbol{r},\downarrow}
 ,
\\
H_f &=& 
-t_f \sum_{\boldsymbol{r},\boldsymbol{r}^{\prime },\sigma} 
f_{\boldsymbol{r},\sigma}^\dagger 
f_{\boldsymbol{r}',\sigma}
-\sum_{\boldsymbol{r}}
 ( 4 t_f -\mu_f) 
  f_{\boldsymbol{r},\uparrow}^\dagger 
  f_{\boldsymbol{r},\uparrow}
\nonumber\\
&&
- \sum_{\boldsymbol{r}} 
( 4 t_f -\mu_f + V_\mathrm{ex}^f)  
f_{\boldsymbol{r},\downarrow}^\dagger
 f_{\boldsymbol{r},\downarrow}
 ,
 \\
  H_{df}
 &=& 
 V_{df} \sum_{\boldsymbol{r},\sigma} 
 \left(
 d_{\boldsymbol{r},\sigma}^\dagger  f_{\boldsymbol{r},\sigma}
 +
  f_{\boldsymbol{r},\sigma}^\dagger  d_{\boldsymbol{r},\sigma}
 \right)
 ,
\end{eqnarray}
where  $d_{\boldsymbol{r} ,\sigma }^{\dagger }(f_{\boldsymbol{r} ,\sigma }^{\dagger }$) is the creation operator, $t_d (t_f)$ is the hopping integral
and  $V_\mathrm{ex}^d (V_\mathrm{ex}^f)$ is the exchange splitting of  $d(f)$ electrons.
The Fermi energy of $d$ and $f$ bands are, respectively, given by $\mu_d=-g_d$ and $\mu_f = 8 t_f + g_f$ , where $g_d (g_f)$ is the energy gap of the $d(f)$ band (see Fig.1).
The third term $H_{df}$ of the Hamiltonian describes the mixing between $d$ and $f$ bands.
It was recognized for a long time that the $d$-$f$ mixing is very important to understand magnetic properties of the Eu chalcogenides.
So we have taken into account the $d$-$f$ mixing term in the Hamiltonian.

The Hamiltonian is diagonalized by the Bogoliubov transformation and the
Bogoliubov-de Gennes (BdG) equation is numerically solved by the recursive
Green function method~\cite{rf:Asano}. We calculate the Matsubara
Green function,
\begin{equation}
\check{G}_{\omega _{n}}(\boldsymbol{r},\boldsymbol{r}^{\prime })=\left(
\begin{array}{cc}
\hat{g}_{\omega _{n}}(\boldsymbol{r},\boldsymbol{r}^{\prime }) & \hat{f}%
_{\omega _{n}}(\boldsymbol{r},\boldsymbol{r}^{\prime }) \\
-\hat{f}_{\omega _{n}}^{\ast }(\boldsymbol{r},\boldsymbol{r}^{\prime }) & -%
\hat{g}_{\omega _{n}}^{\ast }(\boldsymbol{r},\boldsymbol{r}^{\prime })%
\end{array}
\right) , \label{deff}
\end{equation}
where $\omega _{n}=(2n+1)\pi T$ is the Matsubara frequency, $n$ is an
integer number, and $T$ is a temperature. The Josephson current is given by
\begin{equation}
I_J (\phi)=-ietT\sum_{\omega _{n}}\sum_{m=1}^{W}\mathrm{Tr}\left[ \check{G}_{\omega
_{n}}(\boldsymbol{r}^{\prime },\boldsymbol{r})-\check{G}_{\omega _{n}}(%
\boldsymbol{r},\boldsymbol{r}^{\prime })\right]
\end{equation}
with $\boldsymbol{r}^{\prime }=\boldsymbol{r}+\boldsymbol{x}$. 
Throughout this paper we fix the following
parameters: $W=25$, $\mu =2t$, $\Delta _{0}=0.01t$, and $T=0.01T_{c}$ ($T_c$ is the superconductor transition temperature). 
\section{Josephson current through Eu-chalcogenides}

Below, we consider the Josephson transport through the Eu-chalcogenides.
In calculation, we use the following parameters in consideration of EuO~\cite{rf:Steeneken,rf:Sinjukow}: $t_d=1.25$eV, $g=g_d +g_f =1.12$eV, $t_f=0.125$eV, and $V_\mathrm{ex}^d=0.528$eV.

We first discuss the Josephson current through the spin-filtering barrier only [Fig. 3(a)], i.e., the $d$-band.
The phase diagram depending on the strength of $V_\mathrm{ex}^d$ ( $0 \le V_\mathrm{ex}^d/t_d \le 6$) and the thickness of FI $L_F$ is plotted in Fig.~3(b).
In this case, the $\pi$ junction is not formed irrespective of $L_F$, and $V_\mathrm{ex}$.
Therefore, only  the spin-filter effect dose not lead to the $\pi$-junction behaviors.

\begin{figure}[t]
\begin{center}
\includegraphics[width=8.0cm]{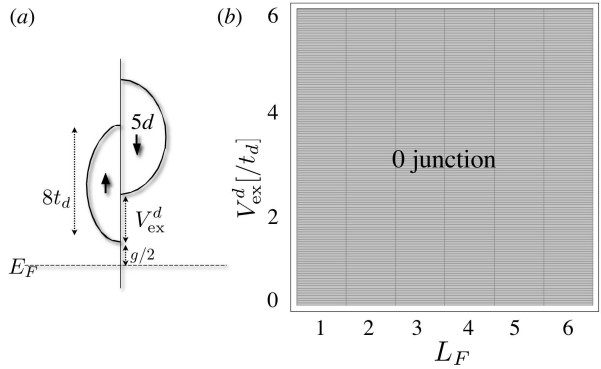}
\end{center}
\caption{(a) The density of states for each spin direction for  the spin-filtering barrier (5$d$ band of Eu).
(b) The phase diagram depending on the strength of $V_\mathrm{ex}$ and $L_F$ for the the spin-filtering barrier. 
In this case, no $\pi$ junction is formed. }
\label{fig3}
\end{figure}
\begin{figure}[h]
\begin{center}
\includegraphics[width=7.5cm]{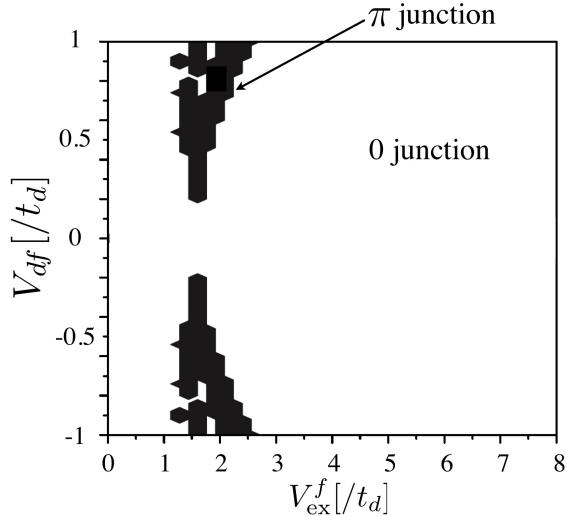}
\end{center}
\caption{The phase diagram depending on the $d$-$f$ hybridization $V_{df}$ and the exchange splitting of the $f$ band $V_\mathrm{ex2}^f$ for the S/Eu-chalcogenides/S Josephson junctions.
The black and white regime correspond to the $\pi$ and 0 junction, respectively. }
\label{fig3}
\end{figure}

Next we consider the Josephson transport through the Eu-chalcogenides with both $d$ and $f$-bands.
In calculation we set $L_F=5$ and systematically change the values of the exchange splitting of $f$ electron $V_\mathrm{ex}^f (=0.0 \sim 10.0$eV) and the $d$-$f$ hybridization $V_{df}(= -1.25\sim~1.25$ eV).
Fig. 4 shows the numerically obtained $0-\pi$ phase diagram.
The $\pi$ junction can be realized at the certain values of $V_{df}$ and $V_\mathrm{ex}^f$.
We found that the $\pi$ junction can be formed if (1) $d$ and $f$ bands are overlapped each other and (2) the $d$-$f$ hybridization $V_{df}$ is strong enough.
More detailed discussion for above results will be given in elsewhere~\cite{rf:Kawabata-pi}.

\section{Summary}
To summarize, we have studied he Josephson effect in S/Eu-chalcogenides/S junction by use of the recursive Green's function method.
The $\pi$ junction behavior is realized if the $d$ and $f$ bands are overlapped and  the $d-f$ hybridization is strong. 
Such Eu-chalcogenides based $\pi$  junctions may becomes a  element in the architecture of "quiet qubit".

\section*{Acknowledgements}

This work was  supported by CREST-JST and a Grant-in-Aid for Scientific Research from the Ministry of Education, Science, Sports and Culture of Japan (Grant No. 19710085).

\newpage
\newpage


\begin{thebibliography}{99}
%
%
\bibitem{rf:Buzdin}
A. I. Buzdin,
Rev. Mod. Phys. {\bf 77}  (2005) 935.
%
%
\bibitem{rf:Golubov}
A. A. Golubov, M. Y. Kupriyanov, E. Il'ichev, 
Rev. Mod. Phys. {\bf 76} (2004) 411.
%
%
\bibitem{rf:Ryanzanov}
V. V. Ryazanov, V. A. Oboznov, A. Y. Rusanov, A. V. Veretennikov, A. A. Golubov, J. Aarts,
Phys. Rev. Lett. {\bf 86} (2001) 2427.
%
%
\bibitem{rf:Kontos}
T. Kontos, M. Aprili, J. Lesueur, F. Gen\^et, B. Stephanidis, R. Boursier,
Phys. Rev. Lett. {\bf 89} (2002) 137007.
%
%
\bibitem{rf:Ioffe}
L B. Ioffe, V. B. Geshkenbein, M. V. Feigel'man, A. L. Fauch\'ere, G. Blatter, 
Nature {\bf 398} (1999) 679.
%
%
\bibitem{rf:Blatter}
G. Blatter, V. B. Geshkenbein, L. B. Ioffe,
Phys. Rev. B {\bf 63} (2001) 174511.
%
%
%
%
\bibitem{rf:Zaikin}
A. D. Zaikin, S. V. Panyukov, 
Sov. Phys. JETP {\bf 62} (1985) 137.
%
%
\bibitem{rf:Tanaka}
Y. Tanaka, S. Kashiwaya,
Physica C {\bf 274} (1997) 357.
%
%
%
\bibitem{rf:Kawabata1}
S. Kawabata, A. A. Golubov,
Physica E {\bf 40} (2007) 386.
%
%
%
\bibitem{rf:Kawabata2}
S. Kawabata, S. Kashiwaya, Y. Asano, Y. Tanaka,
Physica C {\bf 437-438} (2006) 136.
%
%
%
\bibitem{rf:Kawabata3}
S. Kawabata, S. Kashiwaya, Y. Asano, Y. Tanaka, A. A. Golubov,
Phys. Rev. B  {\bf 74} (2006) 180502(R).
%
%
%
\bibitem{rf:Kawabata4}
S. Kawabata, Y. Asano, Y. Tanaka, S. Kashiwaya, A. A. Golubov,
Physica C {\bf 468} (2008) 701.
%
%
%
\bibitem{rf:EuO}
R. Meservey,  P. M. Tedrow,
Phys. Rep. {\bf 238} (1994) 173.
%
%
%
\bibitem{rf:Nagahama}
J. S. Moodera, T. S. Santos, T. Nagahama,
J. Phys. Cond. Mat. {\bf 19} (2007) 165202.
%
%
%
\bibitem{rf:Santos}
T. S. Santos, J. S. Moodera, K. V. Raman, E. Negusse, J. Holroyd, J. Dvorak, M. Liberati, Y. U. Idzerda, E. Arenholz,
Phys. Rev. Lett. {\bf 101} (2008) 147201.
%
%
%
\bibitem{rf:Busch}
G. Busch, P. Junod, P. Wachter,
Phys. Lett. {\bf 12} (1964) 11.
%
%
%
\bibitem{rf:Kawabata-pi}
S. Kawabata, Y. Asano, Y. Tanaka, A. A. Golubov, S. Kashiwaya,
in preparation.
%
%
\bibitem{rf:Kawabata-Asano}
S. Kawabata, Y. Asano,
Int. J. Mod. Phys B 23 (2009) 4320.
%
%
\bibitem{rf:Asano}
Y. Asano, 
Phys. Rev. B  {\bf 63}  (2001) 052512.
%
%
%
\bibitem{rf:Steeneken}
P. G. Steeneken, L. H. Tjeng, I. Elfimov, G. A. Sawatzky, G. Ghiringhelli, N. B. Brookes, D.-J. Huang,
Phys. Rev. Lett.  {\bf 88}  (2001) 047201.
%
%
%
\bibitem{rf:Sinjukow}
P. Sinjukow, W. Nolting, 
Phys. Rev. B  {\bf 69}  (2004) 214432.
%
%
%
\end{thebibliography}
\end{document}